\documentclass[twocolumn,showpacs,preprintnumbers,amsmath,amssymb,prl,superscriptaddress]{revtex4}

\usepackage[]{graphicx}
\usepackage{dcolumn}
\usepackage{bm}
\usepackage{color}

\newcommand{\NA}{N_{\rm A}}
\newcommand{\eSi}{{^{28}\rm Si}}
\newcommand{\Si}{{\rm Si}}

\begin{document}

\preprint{IAC/NAh 28Si NA 1.0}

\title{An accurate determination of the Avogadro constant \\ by counting the atoms in a $\eSi$ crystal}
\author{B.\ Andreas}
\affiliation{PTB -- Physikalisch-Technische Bundesanstalt, Bundesallee 100, 38116 Braunschweig Germany}%
\author{Y.\ Azuma}
\affiliation{NMIJ -- National Metrology Institute of Japan, 1-1-1 Umezono, Tsukuba, Ibaraki 305-8563, Japan}%
\author{G.\ Bartl}
\affiliation{PTB -- Physikalisch-Technische Bundesanstalt, Bundesallee 100, 38116 Braunschweig Germany}%
\author{P.\ Becker}
\affiliation{PTB -- Physikalisch-Technische Bundesanstalt, Bundesallee 100, 38116 Braunschweig Germany}%
\author{H.\ Bettin}
\affiliation{PTB -- Physikalisch-Technische Bundesanstalt, Bundesallee 100, 38116 Braunschweig Germany}%
\author{M.\ Borys}
\affiliation{PTB -- Physikalisch-Technische Bundesanstalt, Bundesallee 100, 38116 Braunschweig Germany}%
\author{I.\ Busch}
\affiliation{PTB -- Physikalisch-Technische Bundesanstalt, Bundesallee 100, 38116 Braunschweig Germany}%
\author{M.\ Gray}
\affiliation{NMI -- National Measurement Institute, Bradfield Road, Lindfield, NSW 2070 Australia}%
\author{P.\ Fuchs}
\affiliation{METAS -- Bundesamt fuer Metrologie, Lindenweg 50, 3003 Bern-Wabern, Switzerland.}%
\author{K.\ Fujii}
\affiliation{NMIJ -- National Metrology Institute of Japan, 1-1-1 Umezono, Tsukuba, Ibaraki 305-8563, Japan}%
\author{H.\ Fujimoto}
\affiliation{NMIJ -- National Metrology Institute of Japan, 1-1-1 Umezono, Tsukuba, Ibaraki 305-8563, Japan}%
\author{E.\ Kessler}
\affiliation{NIST -- Natl.\ Inst.\ of Standards and Technol.\, 100 Bureau Drive, Gatheirsburg, MD 20899, USA}%
\author{M.\ Krumrey}
\affiliation{PTB -- Physikalisch-Technische Bundesanstalt, Bundesallee 100, 38116 Braunschweig Germany}%
\author{U.\ Kuetgens}
\affiliation{PTB -- Physikalisch-Technische Bundesanstalt, Bundesallee 100, 38116 Braunschweig Germany}%
\author{N.\ Kuramoto}
\affiliation{NMIJ -- National Metrology Institute of Japan, 1-1-1 Umezono, Tsukuba, Ibaraki 305-8563, Japan}%
\author{G.\ Mana}
\affiliation{INRIM -- Istituto Nazionale di Ricerca Metrologica, strada delle cacce 91, 10135 Torino Italy}%
\author{P.\ Manson}
\affiliation{NMI -- National Measurement Institute, Bradfield Road, Lindfield, NSW 2070 Australia}%
\author{E.\ Massa}
\affiliation{INRIM -- Istituto Nazionale di Ricerca Metrologica, strada delle cacce 91, 10135 Torino Italy}%
\author{S.\ Mizushima}
\affiliation{NMIJ -- National Metrology Institute of Japan, 1-1-1 Umezono, Tsukuba, Ibaraki 305-8563, Japan}%
\author{A.\ Nicolaus}
\affiliation{PTB -- Physikalisch-Technische Bundesanstalt, Bundesallee 100, 38116 Braunschweig Germany}%
\author{A.\ Picard}
\affiliation{BIPM -- Bureau International des Poids et Mesures, Pavillon de Breteuil, 92312 Sèvres cedex, France}%
\author{A.\ Pramann}
\affiliation{PTB -- Physikalisch-Technische Bundesanstalt, Bundesallee 100, 38116 Braunschweig Germany}%
\author{O.\ Rienitz}
\affiliation{PTB -- Physikalisch-Technische Bundesanstalt, Bundesallee 100, 38116 Braunschweig Germany}%
\author{D.\ Schiel}
\affiliation{PTB -- Physikalisch-Technische Bundesanstalt, Bundesallee 100, 38116 Braunschweig Germany}%
\author{S.\ Valkiers}
\affiliation{IRMM -- Institute for Reference Materials and Measurements, Retieseweg 111, B-2440 Geel, Belgium}%
\author{A.\ Waseda}
\affiliation{NMIJ -- National Metrology Institute of Japan, 1-1-1 Umezono, Tsukuba, Ibaraki 305-8563, Japan}%
\date{\today}

\begin{abstract}
The Avogadro constant links the atomic and the macroscopic properties of matter. Since the molar Planck constant is well known via the measurement of the Rydberg constant, it is also closely related to the Planck constant. In addition, its accurate determination is of paramount importance for a definition of the kilogram in terms of a fundamental constant. We describe a new approach for its determination by "counting" the atoms in 1 kg single-crystal spheres, which are highly enriched with the $^{28}$Si isotope. It enabled isotope dilution mass spectroscopy to determine the molar mass of the silicon crystal with unprecedented accuracy. The value obtained, $\NA = 6.02214084(18)\times 10^{23}$ mol$^{-1}$, is the most accurate input datum for a new definition of the kilogram.
\end{abstract}


\maketitle

Accurate measurements of fundamental constants are a way of testing the limits of our knowledge and technologies. In these tests, the measurement of the Avogadro constant, $\NA$, holds a prominent position because it connects microphysics and macrophysics. In addition, as a new definition of the kilogram most likely will be based on the Planck constant \cite{Mills}, $h$, a determination of $\NA$ is a way to obtain an independent $h$ value via the molar Planck constant, $\NA h$.

While the uncertainty of the mass of the international kilogram-prototype is zero by convention, any new realization will fix an uncertainty to the kilogram. However, since its mass is suspected to have drifted by about of 50 $\mu$g over 100 years, it has been accepted that the relative uncertainty is $2 \times 10^{-8}$, at the maximum. Two experiments have the potential to achieve this goal. One is the watt-balance experiment. It aims at measuring $h$ by the virtual comparison of mechanical and electrical powers \cite{Steiner}. The other experiment aims at determining $\NA$ by counting the atoms in 1 kg nearly perfect single-crystal silicon spheres \cite{Becker:09}. In this method, crystallization acts as a "low noise amplifier" making the lattice parameter accessible to macroscopic measurements, thus avoiding the single atom counting. Silicon is used because it is one of the best known materials and it can be grown into high purity, large, and almost perfect single crystals.

\begin{figure}[b]\centering
\includegraphics[height=7.5cm,angle=-90]{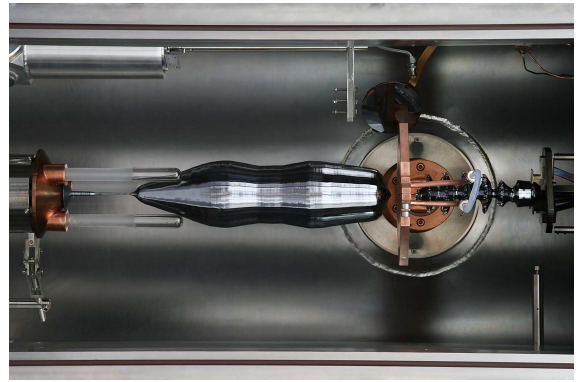}
\caption{The float-zone $\eSi$ crystal. To determine its density, two spheres were manufactured from the two bulges. To determine the lattice parameter, an X-ray interferometer was cut from the material between these spheres.}\label{fig:28Si}
\end{figure}

Since 1998 \cite{CODATA:00}, a relative $1.2\times 10^{-6}$ discrepancy has been observed when comparing the results of these experiments through $\NA h$. It was conjectured that it originated through the difficulty of accurately determining the isotopic composition of a natural Si crystal, a key measurement for $\NA$ determination. To solve this problem, we repeated the measurement by using a silicon crystal highly enriched with the $\eSi$ isotope. In this way, the absolute calibration of the mass spectrometer with the required small uncertainty could be overcome by applying isotope dilution mass spectrometry combined with multicollector inductively coupled plasma mass spectrometry.

The project started in 2004 with the isotope enrichment by centrifugation of SiF$_4$ gas undertaken at the Central Design Bureau of Machine Building in St.\ Petersburg. Subsequently, after conversion of the enriched gas into SiH$_4$, a polycrystal was grown by chemical vapor deposition at the Institute of Chemistry of High-Purity Substances of the Russian Academy of Sciences in Nizhny-Novgorod and, in 2007, the 5 kg $\eSi$ boule shown in Fig.\ \ref{fig:28Si} was grown by the Leibniz-Institut f\"ur Kristallz\"uchtung in Berlin \cite{Becker:06}.

\paragraph{Principle of the measurement.}
Atoms were counted by exploiting their ordered arrangement in the crystal. Provided the crystal and the unit cell volumes are measured and the number of atoms per unit cell is known, the count requires their ratio to be calculated. Hence, $\NA = nM/(\rho_0 a_0^3)$, where $n=8$ is the number of atoms per unit cell, $M/\rho_0$ and $a_0^3$ are the molar and unit-cell volumes, $M$ the molar mass and $\rho_0$ the density. The crystal must be free of imperfections, monoisotopic (or the isotopic composition must be determined), and chemically pure. We selected a spherical crystal-shape to trace back the volume determination to diameter measurements and to make possible an accurate geometrical, chemical, and physical characterization of the surface. Hence, two spheres, AVO28-S5 and AVO28-S8, were taken at 229 mm and 367 mm distances, respectively, from the seed crystal position and shaped as quasi-perfect spheres by the Australian Centre for Precision Optics.

\begin{table}[t]
\caption{\label{table:point}Point-defect concentration (expressed in $10^{15}$ cm$^{-3}$) in the AVO28-S5 and AVO28-S8 spheres and in the X-ray interferometer (XINT).}
\begin{ruledtabular}
\begin{tabular}{llllll}
Defect		&AVO28-S5	&AVO28-S8	&XINT \\
\hline
Carbon		&0.43(9)	&1.85(20)	&0.99(12) \\
Oxygen		&0.21(7)	&0.40(13)	&0.36(4) \\
Boron	    &0.014(5)	&0.04(2)	&0.005(2) \\
Vacancy 	&0.33(10)	&0.33(10)	&0.33(10)

\end{tabular}
\end{ruledtabular}
\end{table}

\paragraph{Imperfections.}
Our boule is dislocation free, it was purified by the float-zone technique, no doping by nitrogen was used, and the pulling speed was chosen in order to reduce the self-interstitial concentration. Unavoidable point-like defects by carbon, oxygen, and boron atoms as well as vacancies strain the crystal and change the sphere mass. To apply the necessary corrections, their concentrations were measured by infrared spectroscopy and positron life time spectroscopy; the results are given in Table \ref{table:point}. Laser scattering tomography excluded voids having diameters greater than the 30 nm detection limit.

\paragraph{Lattice parameter.}
To measure the lattice parameter, an X-ray interferometer was fabricated from the material between the spheres. Next, the mean lattice parameter of each sphere given in Table \ref{table:NA}, $a(\textrm{S}) = (1 + \sum_i \beta_i\Delta N_i)a_0(\textrm{XINT})$, was calculated by taking account of the different contaminations of the spheres and the interferometer. In this equation, S is the sphere AVO28-S5 or -S8, $a_0(\textrm{XINT})$ is the measured value of the interferometer lattice parameter \cite{Massa}, $i$ labels the point defects, $\beta_C = -6.9(5)\times 10^{24}$ cm$^3$, $\beta_O = -4.4(2)\times 10^{24}$ cm$^3$, and $\beta_B = -5.6(2)\times 10^{24}$ cm$^3$ are the strain coefficients \cite{Windisch}, and $\Delta N_i$ is the difference of the defect concentration between the spheres and the interferometer (see Table \ref{table:point}). The lattice parameter of a number of samples, taken from both sides of the spheres, was determined via double-crystal Laue diffractometry. After corrections for the differences in the point-defect concentrations, all the measured values were found to agree within their measurement uncertainties. Lattice parameter topographies, made by using both X-ray phase-contrast imaging and a novel self-referenced X-ray diffractometer, did not show evidence of any intrinsic strain.

\begin{table}
\caption{\label{table:layer}Mass and thickness of the surface layer and mass deficit due to the point defects.}
\begin{ruledtabular}
\begin{tabular}{llll}
 &unit &AVO28-S5 &AVO28-S8 \\
\hline
Surface layer mass	    &$\mu$g	&224(15)	&215(15) \\
Surface layer thickness	&nm	    &2.91(30)	&2.72(28) \\
Mass deficit	        &$\mu$g	&9.3(6)	    &23.7(4)

\end{tabular}
\end{ruledtabular}
\end{table}

\begin{figure}[b]\centering
\includegraphics[width=7.5cm]{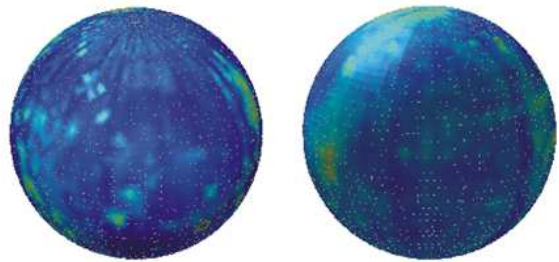}
\caption{Topographic maps of the SiO$_2$ thickness. Left AVO28-S5, right AVO28-S8. The rainbow colours range from 2.0 nm (blue) to 4.5 nm (yellow).}\label{fig:surface}
\end{figure}

\paragraph{Surface.}
Silicon is covered with an oxide surface-layer. X-ray photoelectron spectroscopy, X-ray fluorescence, and NEXAFS (near-edge X-ray absorption fine structure) measurements revealed unexpected surface contamination by copper and nickel silicides. Therefore, to determine the oxide layer mass and thickness given in Table \ref{table:layer}, the sphere surface was modelled, from top to bottom, as follows: a carbonaceous and an adsorbed water layer, a fictive layer of Cu and Ni silicides, and a SiO$_2$ layer \cite{Seah}. The oxide thickness was determined by X-ray fluorescence measurements, where the oxygen K fluorescence intensity from the sphere surface was compared with that from flat samples for which the oxide layer thickness was determined by X-ray reflectometry. The mass of the carbon, copper, and nickel was obtained from X-ray fluorescence measurements. The stoichiometries of the oxide was determined by X-ray photoelectron spectroscopy, which also excluded from consideration an SiO interface. Data for chemisorbed water on silicon were taken from the literature \cite{Mizushima}. Figure \ref{fig:surface} shows the mapping of the surface layer thickness, obtained by spectroscopic ellipsometry with a spatial resolution of 1 mm \cite{Busch}.

\paragraph{Mass.}
The spheres mass given in Table \ref{table:NA} was determined by comparison with the Pt-Ir prototypes of the BIPM, NMIJ, and PTB; the results are in excellent agreement and demonstrate a measurement accuracy of about 5 $\mu$g. Corrections for the surface layers and for the crystal point-defects have to be considered. Owing to point defects, there is the $\Delta m = V\sum_i (m_i-m_{28})N_i$ difference given in Table \ref{table:layer} between the measured mass and the mass of a perfect lattice having a Si atom on each regular site. In this equation, $m_i$ and $m_{28}$ are the masses of, respectively, the $i$-th point defect and of $\eSi$ (a vacancy mass is zero and oxygen is associated to an interstitial lattice site, so that, $m_O$ is the sum of the oxygen and the $\eSi$ masses), $V$ is the sphere volume, and $N_i$ is the concentration of the $i$-th point defect (see Table \ref{table:point}).

\begin{figure}[t]\centering
\includegraphics[width=7.5cm]{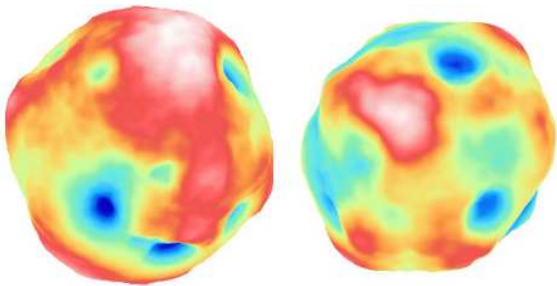}
\caption{Diameter topographies of the silicon spheres. The rainbow colours range from -63 nm (blue) to 37 nm (red). Peak-to-valley distances are 97 nm (AVO28-S5, left) and 89 nm (AVO28-S8, right).}\label{fig:sphere}
\end{figure}

\paragraph{Volume.}
The sphere volumes were determined from diameter measurements carried out by optical interferometry. Two different interferometers were used, both relying on differential measurements \cite{Nicolaus}. Each sphere is placed between the end-mirrors (plane, in one interferometer, spherical, in the other) of a Fizeau cavity and the distances between the mirrors and each sphere, as well as the cavity length, were measured. Since the sphere is almost perfect, its volume is that of a mathematical sphere having the same mean diameter. Hence, a number of diameters were measured and averaged. Figure \ref{fig:sphere} shows the deviations from a constant diameter in orthographic projections. The measured diameters were corrected for phase shifts in beam reflections at the sphere surface, as well for the beam retardation through the surface layer. The final volumes are given in Table \ref{table:NA}.

\begin{figure}\centering
\includegraphics[width=8cm]{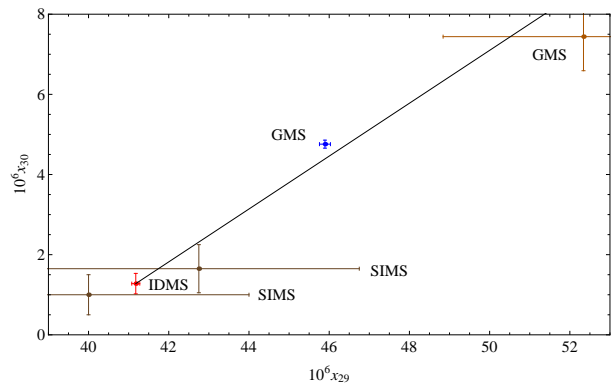}
\caption{Isotope amount fractions of the enriched crystal. Measured isotope amount fractions $x_{29}$ and $x_{30}$ of the $\eSi$ boule as determined by gas mass spectrometry (GMS) by secondary ion mass spectrometry (SIMS), and isotope dilution mass spectrometry  (IDMS). The solid line indicates the composition locus of fictional samples obtained by adding natural silicon to the enriched silicon measured by IDMS. The bars give the standard uncertainties.}\label{fig:molar-mass}
\end{figure}

\paragraph{Molar mass.}
The molar mass is given by $M  = \sum_n M(^n\Si) x_n = \sum_n M(^n\Si) R_{n/m} / \sum_n R_{n/m}$, where $M(^nSi)$ with $n = 28, 29, 30$ is the molar mass of the $^n\Si$ isotope, $x_n$ is the amount-of-substance fraction of $^n\Si$, and $R_{n/m} = x_n / x_m$ are the fraction ratios. The usual way to measure the isotope fractions is by gas mass spectrometry of SiF$_4$. An analysis carried out at the University of Warsaw by graphite-furnace atomic absorption spectroscopy evidenced that the solutions used to convert the approximately 45 mg crystal-samples into the SiF$_4$ gas were contaminated by more than 7 $\mu$g of natural Si. This contamination required a correction greater than $10^{-6}M$. The isotope fractions were measured also at the Institute of Mineral Resources of the Chinese Academy of Science still by gas mass spectrometry, but using a different preparation of the SiF$_4$ gas based on fluorination by BrF$_5$. Also in this case the contamination by natural Si proved to be a problem. Furthermore, the extremely high isotopic enrichment showed to be a big challenge; ion current ratios more than five orders of magnitude larger than the unity had to be measured.

To overcome these difficulties, a novel concept has been developed, which does not require the explicit measurement of the $\eSi$ fraction \cite{Rienitz,Mana}. It is based on isotope dilution mass spectrometry (IDMS) combined with multicollector inductively coupled plasma mass spectrometry - with samples being dissolved in aqueous NaOH. Additionally, the isotopic composition was determined also at the Institute for Physics of Microstructures of the Russian Academy of Sciences by a secondary ion mass spectrometer (SIMS) using a time-of-flight mass analyzer.

Figure \ref{fig:molar-mass} shows the measured $^{29}$Si and $^{30}$Si amount fractions. The highest enrichments are observed by IDMS and SIMS. Continuously adding natural Si to a material having the isotopic composition determined by IDMS, the $^{29}$Si and $^{30}$Si fractions will move along the black line and also pass quite near to the other measurement results. Under this assumption the isotopic compositions determined by IDMS and gas mass spectrometry are consistent with each other. The IDMS data are the most accurate one and have been considered only; the relevant molar mass values are given in Table \ref{table:NA}.

\paragraph{$\NA$ determination.}
The measured values of the quantities necessary to determine $\NA$ are summarized in Table \ref{table:NA}. The $\NA$ determinations based on two spheres differ only by $3(3) \times 10^{-8}\NA$, thus confirming the crystal homogeneity. By averaging these values, the final value of the Avogadro constant is
\begin{equation}\label{NA}
 \NA = 6.02214084(18)\times 10^{23}\; \rm g/mol,	
\end{equation}
with $3.0\times 10^{-8}$ relative uncertainty. The main contributions to the uncertainty budget are given  in Table \ref{table:budget}.

\begin{table}
\caption{\label{table:NA}$\NA$ determination. Lattice parameter, volume, and density are measured at 20.0 $^\circ$C and 0 Pa.}
\begin{ruledtabular}
\begin{tabular}{llrr}
quantity &unit &AVO28-S5 &AVO28-S8 \\
\hline
$a_0$         &pm          & 543.0996234(19)    & 543.0996184(19) \\
$m$           &g           & 1000.087559(15)	& 1000.064540(15) \\
$V$           &cm$^3$      & 431.059060(12)	    & 431.049112(12) \\
$\rho$        &kg/m$^3$    & 2320.070847(74)	& 2320.070990(76) \\
$M$           &g/mol       & 27.97697017(16) 	& 27.97697025(19) \\
\hline\\
$\NA$         &$10^{23}$ mol$^{-1}$  & 6.02214093(21)	&6.02214075(22)
\end{tabular}
\end{ruledtabular}
\end{table}

\begin{table}[b]
\caption{\label{table:budget}Uncertainty budget of the $\NA$ determination.}
\begin{ruledtabular}
\begin{tabular}{lrr}
quantity &relative uncertainty           &contribution \\
         & \multicolumn{1}{c}{$10^{-9}$} & \multicolumn{1}{c}{\%} \\
\hline
Molar mass	&8	&5 \\
Sphere mass	&5	&2 \\
Surface	    &20	&34 \\
Sphere volume	  &22	&41 \\
Lattice parameter &11	&10 \\
Impurities	      &3	&1 \\
Crystal perfection	&9	&7
\end{tabular}
\end{ruledtabular}
\end{table}

\begin{figure}\centering
\includegraphics[width=8cm]{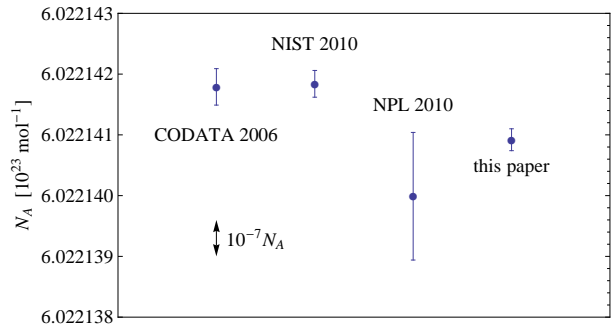}
\caption{Avogadro constant determinations. Comparison between the most accurate values at present available. The bars give the standard uncertainty.}\label{fig:NA}
\end{figure}

\paragraph{Conclusions.}
For the first time precise $h$ values derived from different experiments can be compared. This comparison is a test of the consistency of atomic physics. A parallel experiment, having the purpose of measuring $\NA h$ by absolute nuclear spectroscopy, is aiming at extending this test to nuclear physics \cite{Rainville}. Figure \ref{fig:NA} shows our result compared with those of the two most accurate measurements so far carried out: The watt-balance experiments of the National Institute of Standards and Technology (NIST - USA) \cite{Steiner} and the National Physical Laboratory (NPL - UK) \cite{Robinson}. The values of the Planck constant measured by these experiments were converted into the corresponding $\NA$ values by $\NA h = 3.9903126821(57) \times 10^{-10}$ Js/mol \cite{CODATA:08}.

Our result leads to more consistent numerical values for the fundamental physical constants. It is also a significant step towards demonstrating a successful "mise en pratique" of a kilogram definition based on a fixed $\NA$ or $h$ values. The agreement between the different realizations is not yet as good as it is required to retire the Pt-Ir kilogram prototype, but considering the capabilities already developed and the envisaged improvements it seems to be realistic that the targeted uncertainty may be achieved in the foreseeable future \cite{glaeser}.

We wish to thank A. K. Kaliteevski and his colleagues at the Central Design Bureau of Machine Building and the Institute of Chemistry of High-Purity Substances for their dedication and the punctual delivery of the enriched material, our directors for their advice and financial support, and our colleagues for their daily work. This research received funds from the European Com-munity's 7th Framework Programme ERA-NET Plus (grant 217257) and the International Science and Technology Center (grant 2630).

\end{document}